\documentclass[10pt,reqno]{amsart}

\usepackage{amsfonts}
\usepackage[sans]{dsfont}

\usepackage{amsmath,amssymb,amsthm}
\usepackage[margin=1.0in]{geometry}
\usepackage{enumerate}
\usepackage{color}
\theoremstyle{plain}
\newtheorem{theorem}{Theorem}

\theoremstyle{definition}

\numberwithin{theorem}{section}
\numberwithin{equation}{section}

\newcommand{\R}{\mathbb{R}}
\newcommand{\C}{\mathbb{C}}

\newcommand{\D}{\mathcal{D}}

\newcommand{\cF}{\mathcal{F}}

\newcommand{\cH}{\mathcal{H}}

\newcommand{\cK}{\mathcal{K}}
\def\CO{{\mathcal O}}

\newcommand{\eps}{{\varepsilon}}
\newcommand{\ve}{{\varepsilon}}

\newcommand{\e}{{\epsilon}}
\newcommand{\al}{{\alpha}}

\newcommand{\G}{{\Gamma}}
\newcommand{\del}{{\delta}}
\newcommand{\lam}{{\lambda}}           % Frequently used abbreviations   %

\newcommand{\Om}{\Omega}                %%%%%%%%%%%%%%%%%%%%%%%%%%%%%%%%%%%
\newcommand{\om}{\omega}
\newcommand{\s}{{\sigma}}

\newcommand{\fh}{\mathfrak{h}}

\def\CO{{\mathcal O}}
\renewcommand{\d}{\mathrm{d}}

\def\<{\langle}
\def\>{\rangle}

\newcommand{\im}{\operatorname{Im}}
\newcommand{\re}{\operatorname{Re}}
\newcommand{\supp}{\operatorname{supp}}

\newcommand{\ran}{\rangle}
\newcommand{\lan}{\langle}
\newcommand{\ra}{\rightarrow}

\newcommand{\Ran}{\operatorname{Ran}}
\newcommand{\dist}{\operatorname{dist}}
\newcommand{\p}{{\partial}}

\newcommand{\dive}{\operatorname{div}}

\newcommand{\bfone}{{\bf 1}}
\newcommand{\one}{\mathbf{1}}

\newcommand{\ls}{\lesssim}

\newcommand{\DETAILS}[1]{}
\newcommand{\slim}{\mathop{\text{\rm{s-lim}}}}

\newcommand{\hp}{H_{p}}
\newcommand{\chp}{\mathcal{H}_{p}}

\newcounter{foo}
\Roman{foo}
\begin{document}

\noindent To J\"urg Fr\"ohlich whose vision and ideas shaped\\ 
the non-relativistic quantum electrodynamics\\ 
\,\\
\,\\

\title[Rayleigh scattering in non-relativistic QED]{On Rayleigh scattering in non-relativistic quantum electrodynamics}

\author[J. Faupin]{J{\'e}r{\'e}my Faupin}
\address[J. Faupin]{Institut de Math{\'e}matiques de Bordeaux \\
UMR-CNRS 5251, Universit{\'e} de Bordeaux 1, %\\351 cours de la lib{\'e}ration,
33405 Talence Cedex, France}
\email{jeremy.faupin@math.u-bordeaux1.fr}
\author[I. M. Sigal]{Israel Michael Sigal} %\footnote{ Supported by NSERC Grant No. NA7901.}
\address[I. M. Sigal]{Department of Mathematics \\
University of Toronto, %\\40 St. George Street, Bahen Centre,
Toronto, ON M5S 2E4, Canada}
\email{im.sigal@utoronto.ca}

%\date{October 30, 2012}

\begin{abstract}
In this note we provide details of the proof of the results of \cite{FauSig1} (minimal photon velocity bounds and -- under certain implicit conditions -- asymptotic completeness below the ionization threshold,
i.e. for Rayleigh scattering)  for the standard model of non-relativistic quantum electrodynamics, mentioned in that paper.
\end{abstract}

\maketitle

\section{ Introduction}\label{sec:intro}

In this note we provide details of the proofs of the main results of \cite{FauSig1} to the standard model of non-relativistic quantum electrodynamics in which particles are minimally coupled to the quantized electromagnetic field at energies below the ionization threshold. Recall that in  \cite{FauSig1} we proved several lower bounds on the growth of the distance of the escaping photons/phonons to the particle system. Using some of these results, we proved asymptotic completeness (for Rayleigh scattering) on the states for which the expectation of the photon/phonon number is bounded uniformly in time. However, we provided details only for the phonon case.

\medskip

\noindent \textbf{Model and assumptions.}
We consider the standard model of non-relativistic quantum electrodynamics in which particles are minimally coupled to the quantized electromagnetic field. The state space for this model is given by ${\cH}:=\chp\otimes \cF$, where $\cH_{p}$ is the particle state  space,  $\cF$ is the bosonic Fock space, $\cF\equiv\G(\fh):=\C\oplus_{n=1}^\infty \otimes_s^n \fh$, based on the one-photon space $\fh:=L^2(\R^3, \C^2)$   ($\otimes_s^n$ stands for the symmetrized tensor product of $n$ factors,
 $\C^2$ accounts for the photon polarization).   Its dynamics is generated by the hamiltonian
\begin{equation}\label{Hqed}
H := \sum_{j=1}^n \frac{1}{2 m_j} \big( - i \nabla_{x_j} - \kappa_{j} A_\xi ( x_j ) \big)^2 + U (x) + H_f .
\end{equation}
Here, $m_j$ and $x_j$, $j=1, \ldots , n$, are the (`bare') particle masses and the particle positions, $U(x)$, $x = ( x_1 , \ldots , x_n )$, is the total potential affecting the particles, and $\kappa_{j}$ are coupling constants related to the particle charges. Moreover,
 $A_\xi : = \check \xi *A$, where  $\xi$ is an \textit{ultraviolet cut-off} satisfying e.g. $|\p^m \xi (k)|\ls \lan k\ran^{-3}$,  $|m|=0,1, 2$, and $A(y)$ is the \textit{quantized vector potential} in the Coulomb gauge ($\dive A(y)=0$), describing the quantized electromagnetic field and given by
\begin{equation*} %\label{a73}
A_\xi(y) := \sum_{\lambda=1,2} \int\frac{\d k}{\sqrt{2 \omega ( k )}}  \xi (k) \varepsilon_\lambda (k) \big( e^{i k \cdot y} a_\lambda (k) + e^{- i k \cdot y} a_\lambda^* (k) \big) .
\end{equation*}
Here, $\omega ( k ) = \vert k \vert$ denotes the photon dispersion relation ($k$ is the photon wave vector), $\lambda$ is the polarization, and $a_\lambda(k)$ and $a_\lambda^*(k)$ are photon annihilation and creation operators acting on the Fock space $\cF$ (see  Supplement I for the definition). The operator $H_{f}$ is the quantum hamiltonian of the quantized electromagnetic field, describing the dynamics of the latter, given by  $H_f := \d \Gamma( \om )$, where $\d\G(\tau)$ denotes the lifting of a one-photon operator $\tau$ to the photon Fock space, $\d \Gamma ( \tau )\vert_{ \C } = 0$ for $n=0$ and, for $n\ge 1$,
 \begin{equation}\label{dG}
\d \Gamma( \tau )_{\vert \otimes_s^n \fh} := \sum_{j=1}^{n} \underbrace{1 \otimes \cdots \otimes 1 }_{j-1} \otimes \tau \otimes \underbrace{1 \otimes \cdots \otimes 1 }_{n - j} .
\end{equation}
(See Supplement I for definitions and for the expression of $\d \Gamma ( \tau )$ in terms of the creation and annihilation operators. Here and in what follows, the integrals without indication of the domain of integration are taken over entire $\mathbb{R}^3$.)

This model  goes back to the early days of quantum mechanics  (it appears in the review \cite{Fe32_01} as a well-known model and is elaborated in an important way in \cite{PaFi}); its rigorous analysis was pioneered in \cite{Fr1, Fr2} (see \cite{Sig2, Sp2} for extensive references).

\medskip

 We assume that $U(x) \in L^2_{ \mathrm{loc} }( \mathbb{R}^{3n} )$ and is either confining or relatively bounded with relative bound $0$ w.r.t. $-\Delta_x$, so that the particle hamiltonian $H_p := -\sum_{j=1}^n \frac{1}{2 m_j} \Delta_{x_j}  + U (x) $, and therefore the total hamiltonian $H$, are self-adjoint. Hence $H$ generates the dynamics through the Schr\"odinger equation,
\begin{equation}\label{SE}
 i\p_t\psi_t =H\psi_t.
 \end{equation}
As initial conditions, $\psi_0$, we consider states below the ionization threshold $\Sigma$ (i.e. $\psi_0$ in the range of the spectral projection $E_{(-\infty,\Sigma)}(H)$), which is the (largest) energy $\Sigma > \inf \sigma( H_p ) \ge \inf\sigma( H)$, below which the particle system is well localized:
 \begin{equation}\label{exp-bnd}
\|\lan p\ran^2 e^{\del |x|}f(H)\|\lesssim 1 ,
\end{equation}
for any $0\le \del< \dist(\supp f,\Sigma)$ and any   $f \in \mathrm{C}_0^\infty( (-\infty,\Sigma))$, where $\lan p \ran := (1+|p|^2)^{1/2}$, $p := ( p_1 , \dots , p_n )$, $p_j := - i \nabla_{x_j}$. In other words, states  decay exponentially in the particle coordinates $x$. To guarantee that such an energy exists, we assume that $U(x)$ is such that the particle hamiltonian $H_p$ has discrete eigenvalues below the essential spectrum (\cite{Gr, BFS1, BFS2}). ($\Sigma$  is close to $\inf\s_{\textrm{ess}}(\hp)$ and is directly defined in \cite{Gr}.)

We are thus interested in processes like emission and absorption of radiation, or scattering of photons on an electron bound by an external potential (created e.g. by an infinitely heavy nucleus),  in which the particle system (say, an atom or a molecule) is not being ionized.

\medskip

\noindent \textbf{Problem.} Denote  $ $by $\Phi_{j}$ and $E_{j}$ the $ $eigenfunctions and the corresponding $ $eigenvalues of the $ $hamiltonian $H$ below $\Sigma$, i.e. $E_j < \Sigma$. The following are the key $ $characteristics of the $ $evolution of a physical $ $system, in progressive $ $order  the refined $ $information they provide $ $and in our context:
\begin{itemize}
\item \emph{Local decay}
 stating that some photons are $ $bound to the particle system $ $while others (if any) escape to $ $infinity, i.e. the probability $ $that they $ $occupy any bounded $ $region of the physical $ $space tends to zero, as $t \to \infty$.
\item \emph{Minimal photon velocity bound}
 with speed $c$ stating that,  as $t \to \infty$, with $ $probability $\to 1$, the photons are $ $either bound to $ $the particle system or depart from $ $it with the distance $\ge c' t$, for any $c'<c$.

\noindent Similarly, if the probability that at least $ $one photon is at $ $the distance $\ge c'' t$, $c''> c$, from the $ $particle system $ $vanishes, as $ t \to \infty $, we say that the $ $evolution satisfies the \emph{maximal photon velocity bound}  with speed $c$.
\item  \emph{Asymptotic completeness} 
on the interval $ ( - \infty , \Sigma ) $ stating $ $that, for any $\psi_0 \in  \mathrm{Ran} \, E_{(-\infty, \Sigma)}(H)$  and any $\epsilon > 0$, there are $ $photon wave functions $f_{j\epsilon} \in \mathcal{F}$, with a $ $finite number of photons, s.t.  the solution $\psi_t = e^{ - i t H } \psi_0$ of the Schr\"odinger equation \eqref{SE} satisfies
\begin{equation}\label{AC}
\limsup_{ t \to \infty} \| e^{ - i t H }\psi_0 -\sum_j e^{ - i E_j t} \Phi_{j} \otimes_s e^{ - i H_f t} f_{j\e}\| \le \epsilon.
\end{equation}
In other $ $words, for any $\epsilon>0$ and with probability $\ge 1- \epsilon$, the Schr\"odinger $ $evolution $\psi_t$  approaches asymptotically a $ $superposition of states in which  the particle $ $system with a photon cloud $ $bound to it is in one of $ $its bound states, $\Phi_j$, with additional $ $photons (or possibly none) escaping to infinity with the velocity of light.
\end{itemize}

The reason for $\epsilon>0$ in \eqref{AC} is $ $that, for the state $\Phi_{j} \otimes_s f$ to be well $ $defined, as one would $ $expect, one would have $ $to have a very $ $tight control on the $ $number of photons $ $in $f$, i.e. the number $ $of photons escaping the $ $particle system. (See the remark at the end of Subsection 5.4 of \cite{FauSig1} for a more $ $technical explanation.) For massive $ $bosons, $\epsilon>0$ can be dropped (set to zero), as the $ $number of photons can be bound by $ $the energy cut-off.\footnote{For a discussion of scattering of massless bosons in QFT see \cite{Buch}.}

We define the photon velocity in terms of its space-time (and sometimes phase-space-time) localization. In a quantum theory this is formulated in terms of quantum localization observables and related quantum probabilities. We describe the photon position by  the operator $y := i \nabla_k$ on $L^2( \mathbb{R}^3)$, canonically conjugate to the photon momentum $k$. To test the photon localization, we use the observables $\d \Gamma ( \one_\Om ( y ) )$, where $\one_\Om ( y )$ denotes the characteristic function of a subset $\Om$ of $\R^3$.  We also use the localization observables  $\Gamma ( \one_\Om ( y ) )$,  where   $\G(\chi)$ is the lifting of a one-photon operator $\chi$ (e.g. a smoothed out characteristic function of $y$) to the photon Fock space, defined by
\begin{equation*}%\label{def:Gamma}
\Gamma( \chi )  =  \oplus_{n=0}^\infty (\otimes^n \chi) ,
\end{equation*}
(so that $\Gamma( e^b )=e^{\d\Gamma( b )}$), and then to the space of the total system. The observables  $\d\Gamma ( \one_{\Om} ( y ) ) $  and $\Gamma ( \one_{\Om}  ( y ) )$ have the following natural properties: 
\begin{itemize}
\item $\d \Gamma ( \one_{\Om_1\cup \Om_2}  ( y ) )= \d \Gamma ( \one_{\Om_1} ( y ) )+ \d \Gamma ( \one_{\Om_2} ( y ) )$ and  $\Gamma ( \one_{\Om_1} ( y ) )  \Gamma ( \one_{\Om_2} ( y ) )=0$, for $\Om_1$ and $\Om_2$ disjoint,
\item  $T_g X_\Om ( y )   T_g^{-1} = X_{g^{-1}\Om} ( y ) ,$ where $T_{g} = \Gamma ( \tau_{g} )$, with $\tau_{g} : f(y) \ra f(g^{-1}y)$, and $X_\Om ( y ) $ stands for either $\d \Gamma ( \one_\Om ( y ) )$ or $\Gamma ( \one_\Om ( y ) )$.
\end{itemize}
The  observables $\d\Gamma( \one_\Omega(y) )$ can be interpreted as giving the number of photons in Borel sets $\Om\subset \R^3$. They  are closely related to those used in \cite{FrGrSchl2,Ger,LiLo03_01} (and discussed earlier in \cite{Ma66_01} and \cite{Am})  and are consistent with a theoretical description of the detection of photons (usually via the photoelectric effect, see e.g. \cite{MaWo95_01}). The quantity $\lan \psi, \Gamma ( \one_\Om ( y ) )\psi\ran$ is interpreted as the probability that the photons are in the set $\Om$ in the state $\psi$. This said, we should mention that the subject of photon localization is still far from being settled.\footnote{The issue of $ $localizability of photons $ $is a tricky one and has been $ $intensely discussed in the $ $literature since the $ $1930 and 1932 papers by $ $Landau and Peierls \cite{LaPe} and $ $Pauli \cite{Pa} (see also a $ $review in \cite{Ke}). A set of axioms for $ $localization observables was proposed by $ $Newton and Wigner \cite{NeWi} and Wightman \cite{Wi} and further generalized $ $by Jauch and Piron \cite{JaPi}. Observables describing $ $localization of massless particles, satisfying the $ $Jauch-Piron version of the $ $Wightman axioms, were constructed by Amrein in \cite{Am}.}

The fact that for $ $photons the observables $ $we use depend $ $on the choice of $ $polarization vector fields, $\ve_\lam(k)$, $\lam=1, 2,$\footnote{Since polarization $ $vector fields are not smooth, using them to $ $reduce the results from one set of localization $ $observables to another would limit the $ $possible time decay. However, these vector fields $ $can be avoided by using $ $the approach of \cite{LiLo04_01}.}  is $ $not an impediment here as  our results imply analogous results for e.g.  localization observables of Mandel \cite{Ma66_01}  $ $and of Amrein and $ $Jauch and Piron \cite{Am, JaPi}: $\d\G(f^{\rm man}_\Om)$ and $\d\G(f^{\rm ajp}_\Om)$, where $f^{\rm man}_\Om:= P^\perp\one_\Om ( y )P^\perp$ and $f^{\rm ajp}_\Om :=  \one_\Om ( y ) \cap P^\perp$, respectively,  $ $acting  in the Fock space $ $based on the $ $space $\fh=\mathrm{L}^2_{\textrm{transv}}( \mathbb{R}^3 ; \C^3 ):=\{f\in \mathrm{L}^2( \mathbb{R}^3 ; \C^3): k \cdot f(k) =0\}$ instead of $\fh=\mathrm{L}^2( \mathbb{R}^3 ; \C^2)$. Here  $P^\perp:f(k) \ra  f(k)- |k|^{-2}k\  k\cdot f(k)$ is $ $the orthogonal projection $ $on the transverse $ $vector fields and, for two $ $orthogonal projections $P_1$ and $P_2$, the symbol $P_1\cap P_2$ stand $ $for the orthogonal $ $projection on the largest subspace contained in $\Ran P_1 $ and $\Ran P_2$.

\medskip

We say that the $ $system obeys the $ $\emph{minimal photon velocity bound} if $ $the Schr\"odinger evolution, $\psi_t=e^{-i t H} \psi_0 $, obeys $ $the estimates
\begin{align*}%\label{HP}
\int_1^\infty dt\  t^{-\al'} \big \| \d {\Gamma} (\chi_{\frac{|y|}{ct^{\al}} =1})^{\frac{1}{2}} \psi_t \big \|^2 \lesssim  \| \psi_0 \|_0^2,
\end{align*}
for $ $some norm $\| \psi_0 \|_0$, some $0< \al'\le 1$, and for any $\al>0$ and $c>0$ such that $ $either  $\al <1$, or $\al =1$ and $c<1$. In other words there are no $ $photons which either diffuse or $ $propagate with speed $<1$. Here $\chi_{ \Om} (v)$ denotes a $ $smoothed out characteristic $ $function of the set $\Om$, which is defined at the $ $end of the introduction. The \textit{maximal velocity estimate}, as proven in \cite{BoFaSig}, states $ $that, for any $ c'>1$, and $\gamma < \min ( \frac{1}{2} ( 1-  \frac{1}{ c' } ) , \frac{1}{10} )$,
\begin{equation*}%\label{maxvel-est}
\big\Vert \d \Gamma \big( \chi_{ \frac{|y|}{c' t}  \geq 1  } \big)^{\frac{1}{2}} \psi_t  \big\Vert \lesssim t^{- \gamma} \big\Vert ( \d \Gamma ( \< y \> ) + 1 )^{\frac12}  \psi_0 \big\Vert.
\end{equation*}

Considerable progress has $ $been made in understanding $ $the asymptotic dynamics of non-relativis\-tic particle systems coupled $ $to quantized electromagnetic or phonon $ $field. The local decay property was proven in \cite{BFS2, BFSS, BoFa, CFFS, FGSig1, FGSig2, GGM1,GGM2}, by the combination of the $ $renormalization group  and positive commutator methods. The maximal velocity $ $estimate was proven in \cite{BoFaSig}.

An $ $important breakthrough $ $was achieved recently in \cite{DRK}, where the authors proved $ $relaxation to the ground state and $ $uniform bounds on the number of emitted massless bosons in the spin-boson model.  (Importance of both questions was emphasized earlier by J\"urg Fr\"ohlich.)

In quantum field theory, asymptotic completeness was proven $ $for (a small perturbation of) a solvable model involving a $ $harmonic oscillator (see \cite{Ar, Sp1}), and for models $ $involving massive boson $ $fields (\cite{DerGer2,  FrGrSchl2, FrGrSchl3, FrGrSchl4}). Moreover, \cite{Ger} obtained $ $some important results for $ $massless bosons (the Nelson model) in confined $ $potentials (see below for a more $ $detailed discussion). Motivated by the many-body $ $quantum scattering, \cite{DerGer2, FrGrSchl2, FrGrSchl3, FrGrSchl4, Ger} defined the main $ $notions of scattering theory on Fock $ $spaces, such as wave $ $operators, asymptotic completeness $ $and propagation estimates.

\medskip

\noindent \textbf{Results.} Now we formulate $ $our results.  It is $ $known (see \cite{BFS2, GLL}) that the $ $operator $H$ has a \textit{unique ground state} (denoted here as $\Phi_{\textrm{gs}}$) and that generically (e.g. under the $ $Fermi Golden $ $Rule condition) it has no $ $eigenvalues in the interval $(E_{\textrm{gs}},\Sigma)$, where $E_{\textrm{gs}}$ is the ground state energy (see \cite{BFSS, FGSig1, GGM2}). \textit{We assume} that  this is $ $exactly the case:
\begin{equation}\label{EVcond}
\mbox{{\it Fermi's Golden Rule} (\cite{BFS1,BFS2}) holds.}
\end{equation}
(If the particle system has an infinite number of eigenvalues accumulating to its ionization threshold -- which is the bottom of its essential spectrum -- then, to rule out the eigenvalues in the spectral interval of interest, we should replace $\Sigma$ by $\Sigma -\e$ for some fixed $\e$. This is understood from now on.)

We will consider the following sets of initial conditions
\begin{equation*}
\Upsilon_1 := \big \{ \psi_0 \in  f (H) D(\d\G( \omega^{-1} )^{\frac12}), \mbox{ for some }  f \in \mathrm{C}_0^\infty( (-\infty, \Sigma) ) \big \},
\end{equation*}
and  
\begin{equation*}
\Upsilon_2 := \big \{ \psi_0\in  f (H)  D(\d\G(\lan y\ran)), \mbox{ for some }    f \in \mathrm{C}_0^\infty( ( E_{\mathrm{gs}}, \Sigma) ) \big \}.
\end{equation*}

For $A\ge -C$, let $\| \psi_0 \|_A := \| (A+C +1)^{\frac{1}{2}}\psi_0 \|$. We define $\nu(\rho)\ge 0$ as the smallest real number satisfying the inequality
 \begin{equation}\label{dGk-bnd}
\langle \psi_t , \d \Gamma( \om^{\rho} ) \psi_t \rangle \lesssim  t^{\nu(\rho)}\| \psi_0 \|^2_\rho,
\end{equation}
for any $\psi_0 \in \mathrm{Ran} \, E_{ (-\infty , \Sigma) }( H )$, where $\|\psi\|_\rho := \| \psi \|_{\d\G(\omega^{\rho})}$. It was shown in  \cite{BoFaSig} (see also (A.1) of Appendix A of \cite{FauSig1}) that 
\begin{equation*}%\label{nurho}
\nu(\rho)\le \frac{2(1-\rho)}{5}
\end{equation*}
(this generalizes an earlier bound due to \cite{Ger}). Also, the bound $ \|  \psi_t \|_{H_f} \lesssim \| \psi_0 \|_H $ shows that \eqref{dGk-bnd} holds for $\rho=1$ with $\nu(1)= 0$. With $\nu(\rho)$ defined by \eqref{dGk-bnd}, we prove the following two results.
\begin{theorem}[Minimal photon velocity bound]\label{thm:mve1y}
Let either  $\beta =1$ and $c<1$, or \begin{equation}\label{beta-cond_0}
 \frac{1}{6}( 5+ \nu(-1)-\nu(0)) < \beta < 1.
\end{equation}
Then for  any initial condition $\psi_0 \in \Upsilon_{1}$, the Schr\"odinger evolution, $\psi_t$, satisfies, for any $a>1$, the following estimate
\begin{align}\label{mve1y_0}
\int_1^\infty dt\  t^{-\beta-a\nu(0)} \big \| \d {\Gamma} ( \chi_{\frac{\mid y\mid}{ct^{\beta}} =1})^{\frac{1}{2}} \psi_t \big \|^2
 \lesssim   \| \psi_0 \|_{-1}^2.
\end{align}
\end{theorem}

\medskip

\begin{theorem}[Weak minimal photon escape velocity estimate]\label{thm:mve2y}
Let the coupling constants $\kappa_j$ be sufficiently small. Assume \eqref{EVcond} and $\nu(-1) < \alpha < 1- \nu(0)$. Then  for any initial condition $\psi_0 \in \Upsilon_{2}$, the Schr\"odinger evolution, $\psi_t$, satisfies the estimate 
\begin{equation}\label{mve2y}
\big \| \Gamma (\chi_{\frac{|y|}{c' t^\alpha} \le 1}) \psi_t \big \|  \lesssim t^{-\gamma} \big ( \| \psi_0 \|_{\d \Gamma ( \lan y \ran ) }^2 +\| \psi_0 \|^2_{ \d \Gamma ( \tilde b )^2} \big ) ,
\end{equation}
where $\gamma < \frac12 \min ( 1-\alpha - \nu(0) , \frac12 (\alpha - \nu(0)-\nu(-1)) )$  and $\tilde b := \frac{1}{2} ( k \cdot y + y \cdot k )$.
\end{theorem}

\medskip

\noindent \textbf{Remarks.}

1) The estimates \eqref{mve1y_0} and \eqref{mve2y} are sharp if $\nu(0)=0$ and $\nu(-1)=0$.  

2) The estimate \eqref{mve2y} states that, as $t\ra \infty$, with probability $\ra 1$, either all photons are attached to the particle system in the ground state, or at least one photon departs the particle system with the distance growing at least as $\CO(t^\al)$. 

3) The norm on the r.h.s. of \eqref{mve2y} above is different (weaker) than the one given in \cite{FauSig1}, however the proof can be easily upgraded to the weaker norm indicated, by omitting the unnecessary estimate  $\| \psi_0 \|^2_{ \d\Gamma(\tilde b)^2 } \lesssim \| \psi_0 \|^2_{ \d \Gamma( \lan y \ran )^2 }$ for $\psi_0\in D(H)$.

\medskip

Let $N:=\d \Gamma( \one )$ be the photon number operator and  $K:=\d\Gamma( \omega^{-1} )$ be the photon  low momentum number operator. Our next result is
\begin{theorem}[Asymptotic Completeness]\label{thm:ac}
Let the coupling constants $\kappa_j$ be  sufficiently small.  Assume \eqref{EVcond} and suppose that, uniformly in $t \in [0 , \infty)$, either
\begin{equation}\label{npb-unif}
\| N^{1/2} \psi_t \| \lesssim \| N^{1/2} \psi_0 \| + \| \psi_0 \| ,
\end{equation}
for any   $\psi_0 \in f (H) D(N^{1/2})$, with $f \in \mathrm{C}_0^\infty( ( E_{\mathrm{gs}}, \Sigma) )$, or
\begin{equation}\label{npb-unif2}
\| K^{\frac12} \psi_t \| \ls 1, 
\end{equation}
for any  $\psi_0 \in \mathcal{D}$,  where $\mathcal{D}$ is such that $\mathcal{D} \cap D ( \d\G( \omega^{-1/2}Ê\langle y \rangle \omega^{-1/2} )^{\frac12} )$ is dense in $\mathrm{Ran} \, E_{(-\infty , \Sigma) }(H)$. Then the asymptotic completeness holds on $\mathrm{Ran} \, E_{(-\infty , \Sigma) }(H)$.
\end{theorem}

\medskip

The advantage $ $of $ $Assumption \eqref{npb-unif2} is that the uniform bound on $K = \d\Gamma( \omega^{-1} )$ is required $ $to hold only for an \emph{arbitrary} dense set of $ $initial $ $states and, as a result,  can be verified for the $ $massless spin-boson model $ $by modifying slightly $ $the proof of \cite{DRK} (see \cite{FauSig2}). Hence the asymptotic completeness for the massless spin-boson model $ $holds with no implicit conditions.

As we see $ $from the results above, the uniform bounds, \eqref{npb-unif} or \eqref{npb-unif2}, on the number of photons $ $emerge as the remaining stumbling blocks to $ $proving the asymptotic completeness without qualifications. The difficulty in proving these $ $bounds for massless fields is due to the same infrared  $ $problem which pervades this field $ $and which $ $was successfully $ $tackled in other central issues, $ $such as the theory of ground states and resonances (see \cite{bach, Sig2} for  reviews), the local $ $decay and the $ $maximal velocity bound.

For $ $massive $ $bosons,  the inequality \eqref{npb-unif} (as well as \eqref{dGk-bnd}, with $\nu(0)=0$) is $ $easily proven and $ $the proof below $ $simplifies considerably as $ $well. In this case, the result is unconditional. It was first proven in \cite{DerGer2} for $ $models with $ $confined particles, and in \cite{FrGrSchl2} for Rayleigh scattering.

\medskip

\noindent \textbf{Comparison with earlier results.}
For models involving massive bosons fields, some minimal velocity estimates are proven in \cite{DerGer2}. For massless bosons, Theorems \ref{thm:mve1y} and \ref{thm:mve2y} seem to be new. As was mentioned above,  asymptotic completeness was proven for (a small perturbation of) a solvable model involving a harmonic oscillator (see \cite{Ar, Sp1}), and, for models involving massive boson fields, in \cite{DerGer2} for confined systems, in \cite{FrGrSchl2} below the ionization threshold for non-confined systems, and in \cite{FrGrSchl3} for Compton scattering.

The paper \cite{Ger} treats the Nelson model given by  the state space ${\cH}:=\chp\otimes \cF$, where $\cH_{p}:= L^2(\R^3)$,  $\cF$ is the bosonic Fock space based on the one-phonon space  $\fh :=L^2(\R^3, \C)$, and
 the hamiltonian
\begin{equation}\label{H}
H:=\hp + H_f + I(g),
\end{equation}
acting on ${\cH}$.
Here $\hp:=-\frac{1}{2 m} \Delta_x + V(x),$  with $V(x)$ growing at infinity as $V(x)\ge c_0|x|^{2\al} -c_1$, $c_0 > 0$, $\al>0$,
acting on $L^2(\R^3)$,  $H_f=\d \Gamma( \om )$, where $ \omega(k)= |k|$ and $I(g)$ is given by
\begin{equation}\label{I}
I(g):= \int (g^* (k)\otimes a(k)+g (k)\otimes a^*(k))dk,
\end{equation}
where $a^*(k)$ and  $a(k)$ are the phonon creation and annihilation operators acting on $\cF$, with abstract conditions on the coupling function $g$ (allowing a coupling function of the form $ g (k) = |k|^\mu \xi (k) e^{ikx}$, where $\xi (k)$ is the ultraviolet cut-off, with various conditions on $\mu$ depending on the results involved). In this case, in particular, the ionization threshold $\Sigma$ is equal to $\infty$.

We reproduce the main results of \cite{Ger} (Theorems 12.4, 12.5 and 13.3), which are coached in different terms than ours and present another important view of the subject.  Let $f, f_0\in \mathrm{C}^\infty(\R)$ such that $0\le f, f_0\le 1$, $f'\ge 0$, $f=0$ for $s \le \alpha_0$, $f=1$ for $s \ge \alpha_1$, $f_0' \le 0$, $f_0 =1$ for $s \le \al_1$, $f_0=0$ for $s \ge \al_2$, with $0<\al_0<\al_1<\al_2$. Let $P^+_c := \inf_{ c < c' } \hat P^+_{c'}$, with $\hat P^+_{c'} := \slim_{\e\ra 0} \e^{-1}\hat R^+_c(\e^{-1})$,  $\hat R^+_c(\e^{-1}):=\slim_{t\ra \infty} e^{itH}(B_{ct}+\lam)^{-1}e^{-itH}$, $B_{ct}:=\d\G(b_{ct})$, $b_{ct}:=f (\frac{|y|-ct}{t^\rho})$ and  $\G_{c'}^+(f_0):=\slim_{t\ra\infty} e^{itH} \G(f_{0, c', t})e^{-itH} $, where $f_{0, c', t}:=f_0(\frac{|y|-c't}{t^\rho})$.  Then Proposition 12.2 and Theorem 12.3  state that the operators $P^+_c$ exist provided $\rho>\frac{1}{\mu+1}$, are independent of the choice of $f$, and are orthogonal projections commuting with $H$. Furthermore, let $\cK^+:= \{ \Phi \in \mathcal{H} , a_\pm(h) \Phi = 0 , \, \forall h \in \mathfrak{h}\}$ (called in \cite{Ger} the set of asymptotic vacua), where (formally) $a_\pm(h):=\slim_{t\ra \pm\infty} e^{itH} a(e^{-it\om}h)e^{-itH} $ and $ \cH^+_c:= \Ran P^+_c$ (the spaces containing states with only a finite number of photons in the region $\{ |y| \ge c' t \}$ as $t \to \infty$, for all $c' > c$). Assuming $\al>1$ and $\mu>0$, Theorems 12.4 and  12.5 state that the operator $\G_{c'}^+(f_0)$ exists and is equal to the orthogonal projection on the space $\cK_c^+:=\cK^+\cap \cH^+_c$, provided $0<c<c'<1$ and $\rho>\frac{1}{\mu+1}$. Assuming in addition that the Mourre estimate $\one_\Delta (H)[H, i B] \one_\Delta (H)\ge c_0\one_\Delta (H) +R$ holds on an open interval $\Delta\subset \R$, with the conjugate operator $B:=\d\Gamma(\tilde b)$, $c_0>0$ and $R\in \cH$ compact, then for $0< c<c(\Delta, c_0)$, one has $\one_\Delta (H)\cK_c^+=\one_\Delta (H)\cH_{\mathrm{pp}}$, where $\cH_{\mathrm{pp}}$ is the pure point spectrum eigenspace of $H$. (The latter property is called in \cite{Ger} geometric asymptotic completeness. Combining results of \cite{BFS2, BFSS, FGSig1}  one can probably prove a Mourre estimate,  with $ B$ as conjugate operator, in any spectral interval above $E_{\mathrm{gs}}$ and below  $\Sigma$ and for the coupling function $g$ given by $ g (k) = |k|^\mu \xi (k) e^{ikx}$, with $\mu \ge 1/2$.)

Our approach is similar to the one of \cite{Ger} in as much as it also originates in ideas of the quantum many-body scattering theory. At this the similarities end.

\medskip

\noindent \textbf{Approach and organization of the paper.}
In \cite{FauSig1} we gave the details of the proofs of similar results for the hamiltonian \eqref{H}--\eqref{I}, with the coupling operators $g (k)$ satisfying
 \begin{equation}\label{g-est}
\|\eta^{|\al|}\p^\al g (k)\|_{\chp} \lesssim |k|^{\mu-|\al|} \lan k\ran^{-2-\mu},\quad |\al| \le 2,
\end{equation}
where  $\mu> - 1/2$ or $\mu>0$,  and $\eta$ is an estimating operator on the particle space $\chp$ (a bounded, positive operator with unbounded inverse), satisfying
\begin{equation*}%\label{eta-bnd}
\|\eta^{-n}f(H)\|\lesssim 1 ,
\end{equation*}
for any $n=1, 2$ and  $f \in \mathrm{C}_0^\infty( (-\infty,\Sigma))$.  As was mentioned in \cite{FauSig1}, to extend these proofs to the hamiltonian  \eqref{Hqed}, we use,  similarly to \cite{BoFaSig}, the generalized Pauli-Fierz canonical transform to map the hamiltonian \eqref{Hqed} to a hamiltonian $\tilde H$ with milder infrared behaviour (Section \ref{sec:gener-PF}). Then we extend readily techniques of \cite{FauSig1} to $\tilde H$ and relate the results for  $\tilde H$ with those for \eqref{Hqed} (Section \ref{sec:qed-pfs}). We also include Supplement I defining and discussing the creation and annihilation operators and presenting some standard commutation relations.

 \medskip

\noindent \textbf{Notations.} For functions $A$ and $B$, we will use the notation $A\lesssim B$ signifying that $A\le C B$ for some absolute (numerical) constant $0<C < \infty$.
The symbol $E_\Delta $ stands for the characteristic function of a set $\Delta$, while $\chi_{ \cdot \le 1 }$ denotes a smoothed out characteristic function of the interval $(-\infty , 1]$, that is it is in $\mathrm{C}^\infty( \mathbb{R} )$,  is non-decreasing, and $ = 1$ if $x \le 1/2$ and $= 0$ if $x \ge 1$. Moreover, $\chi_{ \cdot \ge 1 } := \one - \chi_{ \cdot \le 1 }$ and $\chi_{ \cdot = 1 }$ stands for the derivative of $\chi_{ \cdot \ge 1 }$. Given a self-adjoint operator $a$ and a real number $\al$, we write $\chi_{ a \le \al } := \chi_{ \frac{ a }{ \al } \le 1 }$, and likewise for $\chi_{ a \ge \al }$. Finally, $D(A)$ denotes the domain of an operator $A$.

\medskip

\section{Generalized Pauli--Fierz transformation}\label{sec:gener-PF}
We consider  for simplicity a single negatively charged particle in an external potential. Then,  absorbing the absolute value of the particle charge into the vector potential $A(x) := A_\xi (x)$ and choosing units such that the electron mass is $m=1/2$,  the hamiltonian \eqref{Hqed} becomes
\begin{equation}\label{Hqed-one-b}
H = \big( p + A (x) \big)^2 + H_f + U (x).
\end{equation}

The coupling function  $g^{\rm qed}_{x} ( k , \lambda ) := |k|^{-1/2}Ê \xi(k) \varepsilon_\lambda(k) e^{ i k \cdot x }$ in \eqref{Hqed-one-b}  is more singular in the infrared than is allowed by our techniques ($\mu>0$). To go around this problem we use the (unitary) generalized Pauli--Fierz transformation (see \cite{Sig1})
\begin{equation}\label{gen-PF}
\mathcal{U} := e^{  i \Phi ( q_{x} )} ,
\end{equation}
to pass from $H$, given in \eqref{Hqed-one-b},  to the new  hamiltonian $\tilde{H} := \mathcal{U} H \mathcal{U}^*$,  where $\Phi( h ) $  is  the operator-valued  field, $\Phi( h ) := \frac{1}{ \sqrt{2} } ( a^*(h) + a(h) )$ and the function $q_{x} (k,\lambda)$ is defined as follows. Let $\varphi \in \mathrm{C}^\infty ( \mathbb{R} ; \mathbb{R} )$ be a non-decreasing function such that $\varphi(r) = r$ if $|r| \leq 1/2$ and $| \varphi(r) | = 1$ if $|r| \geq 1$. For $0 < \nu < 1 / 2$, we define 
\begin{equation}\label{<}
q_{x} (k,\lambda) :=  \frac{\xi(k) }{ |k|^{\frac{1}{2} + \nu } } \varphi( |k|^\nu \varepsilon_\lambda(k) \cdot x ).
\end{equation}
We note that the definition of $\Phi( h )$ gives $A( x )  = \Phi ( g^{\rm qed}_{x} )$. Using \eqref{Phi4} and \eqref{Phi5},  we compute 
\begin{equation}\label{tildeH}
\tilde{H}  = \big( p + \tilde{A} (x) \big)^2 + E(x) + H_f + V (x) ,
\end{equation}
where
\begin{align*}%\label{tildeAEV} 
\begin{cases}
& \tilde{A} (x) := \Phi( \tilde{g}_{x} ), \quad \tilde{g}_{x} ( k,\lambda) := g^{\rm qed}_{x} ( k,\lambda) - \nabla_x q_{x} ( k,\lambda), \\
&E(x) := \Phi( e_{x} ), \quad e_{x} ( k,\lambda) := i |k| q_{x} ( k,\lambda),  \\ 
& V (x) :=U(x) + \frac{1}{2} \sum_{\lambda=1,2} \int_{ \mathbb{R}^3 } | k | | q_{x} ( k , \lambda ) |^2 \d k. 
\end{cases}
\end{align*}
The operator $\tilde{H}$ is self-adjoint with domain $\D( \tilde{H} ) = \D( H ) = \D ( p^2 + H_f )$ (see \cite{HaHe08_01,Hi02_01}).

Now,  the coupling functions (form factors)  $\tilde{g}_{x} ( k , \lambda )$ and $e_{x} ( k , \lambda )$ in the transformed hamiltonian, $\tilde H$, satisfy the estimates that are better behaved in the infrared (\cite{BoFaSig}):
\begin{align*}
& |\partial_{k}^m  \tilde{g}_x(k,\lambda) | \lesssim  \lan k\ran^{-3}  \vert k \vert^{\frac{1}{2} - \vert m \vert} \< x \>^{\frac{1}{\nu} + \vert m \vert} ,  \\
& |\partial_{k}^m  e_x(k,\lambda) | \lesssim \lan k\ran^{-3}  |k|^{\frac{1}{2} - \vert m \vert} \< x \>^{1 + \vert m \vert}.  
\end{align*}
We see that the new hamiltonian \eqref{tildeH} is of the form \eqref{H}, $\widetilde{H}  = H_p + H_f + \tilde I ( g)$,  with  $\hp:=-\Delta + V(x),$ $H_f = \d\G(\om)$  and
\begin{align*}%\label{tildeIqed}
\tilde I(g) &:=\sum_{ij } \iint  \, d \underline{k}_{(i)} d\underline{k}'_{(j)} g_{ij } ( \underline{k}_{(i)}, \underline{k}'_{( j)}) \otimes  a^*(\underline{k}_{(i)})  a ( \underline{k}'_{(j)}),
\end{align*}
where the summation in $i, j$ ranges over the set $i, j \ge 0 , 1 \le i+j \le 2$, $\underline{k}_{(p)}:= ( \underline{k}_1, \dots, \underline{k}_p)$, $\underline{k}_{j} :=  (k_{j},\lambda_{j})$, $\int  d\underline{k}_{(p)} := \prod_1^p \sum_{\lambda_j} \int dk_j$, $a^\#(\underline{k}_{(p)}) :=\prod_1^p a^\#(\underline{k}_{j}) $ if $p\ge 1$ and $=\one$, if $p=0$, $a^\#(\underline{k}_{j}):= a_{\lambda_{j}}^\#(k_{j})$,  and $g:=(g_{ij })$. The coupling operators, $g_{ij }=g_{ij } (\underline{k}_{(i)} , \underline{k}_{(j)}) $ obey
 \begin{equation*}%\label{g-sym} 
 g_{ij }( \underline k_{(i)}, \underline k'_{( j)}) = g_{ji }^* (\underline k'_{( j)}, \underline k_{(i)}),
 \end{equation*}
and satisfy the estimates 
\begin{equation}\label{g-est-qed}
\|\lan p\ran^{i+j-2}\lan x\ran^{-|\al| - \frac{1}{\nu} }\p^\al  g_{ij } ( \underline{k}_{(i+j)} ) \|_{\chp} \lesssim \sum_{m=1}^{i+j} \prod_{\ell=1}^{i+j} (|k_\ell|^{\mu} \lan k_\ell\ran^{-3})|k_m|^{-|\al|} , 
\end{equation}
with $\mu=1/2 , \  | \alpha | \le 2 $, and $1\le i +j\le 2$.

The bound \eqref{exp-bnd} holds for both \eqref{Hqed-one-b}  and \eqref{tildeH}.  With \eqref{exp-bnd}  and \eqref{g-est-qed},  it is easy to extend Theorems \ref{thm:mve1y} and \ref{thm:mve2y} to the operator $\tilde H$ and then translate the result to \eqref{Hqed-one-b}. Similarly one proceeds with Theorem \ref{thm:ac}.  We explain all this in Section \ref{sec:qed-pfs}.

\bigskip

\section{ Proof of Theorems \ref{thm:mve1y}--\ref{thm:ac}}\label{sec:qed-pfs}

\subsection{ Proof of Theorems \ref{thm:mve1y}--\ref{thm:mve2y}}\label{sec:mv-qed}

Using \eqref{g-est-qed},  it is not difficult to extend Theorems \ref{thm:mve1y} and \ref{thm:mve2y} to the operator $\tilde H$ (cf. \cite{BoFaSig}, where one can find many needed, or similar,  estimates).  Then  we translate  Theorems \ref{thm:mve1y} and \ref{thm:mve2y} from $\tilde H$ to the QED hamilonian \eqref{Hqed-one-b}, by using the following estimates (\cite{BoFaSig})
\begin{align} 
&\Big\Vert \d \Gamma ( \chi_1 (  v ) )^{\frac{1}{2}} \psi \Big\Vert^2 \lesssim \Big\< \mathcal{U} \psi , \d \Gamma ( \chi_1  (  v  ) ) \mathcal{U}\psi \Big\>+ t^{-\al d} \Vert \psi \Vert^2, \label{g3-qed} \\
&  \Big\Vert \Gamma ( \chi_2 (  v ) )^{\frac{1}{2}} \psi \Big\Vert^2 \lesssim \Big\< \mathcal{U} \psi ,  \Gamma ( \chi_2  (  v  ) ) \mathcal{U}\psi \Big\>+ t^{-\al d} \Vert \psi \Vert^2, \label{g3-qed2}
\end{align}
where  $v:=\frac{ y}{ct^{\al}}$,  valid for any functions $ \chi_1 ( v ) $ and $ \chi_2 ( v ) $  supported in $\{ \vert v\vert \le \e\}$ and  $\{ \vert v\vert \ge \e\}$, respectively, for some $\e>0$,  for any $\psi \in  f (H) \cH$, with $f \in \mathrm{C}_0^\infty( (-\infty, \Sigma) )$,   and  for $0 \leq d < 1/2$. \eqref{g3-qed} follows from estimates of Section 2 of \cite{BoFaSig} and \eqref{g3-qed2} can be obtained similarly  (see \eqref{Phi5} and \eqref{Phi6}).  Using these estimates for $\psi_t=e^{-i t H} \psi_0 $, with  an initial condition $\psi_0 $ in either $\Upsilon_{1}$ or $\Upsilon_{2}$, together with $\mathcal{U} e^{-i t H} \psi_0 = e^{-i t \tilde H} \mathcal{U} \psi_0 $, and applying Theorems \ref{thm:mve1y} and \ref{thm:mve2y}  for $\tilde H$ to the first terms on the r.h.s., we see that, to obtain  Theorems \ref{thm:mve1y} and \ref{thm:mve2y}  for the hamiltonian \eqref{Hqed-one-b}, we need, in addition, the estimates
\begin{align} 
& \Big\< \psi , \mathcal{U}^*  \d \Gamma ( \om^{- 1} ) \mathcal{U} \psi\Big\> \lesssim \big\< \psi , \big( \d \Gamma ( \om^{- 1} ) + 1 \big) \psi \big\> , \label{1overk-est-qed} \\
& \Big\< \psi , \mathcal{U}^*  \d \Gamma ( \lan y \ran) \mathcal{U} \psi\Big\> \lesssim \big\< \psi , \big( \d \Gamma ( \lan y \ran ) + \< x \>^{2}  \big) \psi \big\> ,  \label{dGy-est-qed} \\
& \big \| \mathcal{U}^*  \d \Gamma (\tilde b) \mathcal{U} \psi \big \| \lesssim  \big \|\big( \d \Gamma ( \tilde b ) + \< x \>^{2}  \big) \psi \big \| ,  \label{dGtildeb-est-qed}
\end{align}
where, recall, $\tilde b = \frac{1}{2} ( k \cdot y + y \cdot k )$.

To prove \eqref{1overk-est-qed}, we see that, by \eqref{Phi5}, we have
\begin{align*} %\label{U*dGU1} 
\mathcal{U}^* \d \Gamma ( \om^{-1} ) \mathcal{U} = e^{i \Phi (q_x)} \d \Gamma ( \om^{-1} ) e^{- i \Phi (q_x)} = \d \Gamma ( \om^{-1} ) - \Phi ( i \om^{-1} q_x ) + \frac{1}{2}  \|\om^{-1/2} q_x \|_{\mathfrak{h}}^2. 
\end{align*}
(Since $\om^{-1} q_x \notin \mathfrak{h}$, the field operator $\Phi( i \om^{-1} q_x )$ is not well-defined and therefore this formula should be modified by introducing,  for instance, an infrared cutoff parameter $\sigma$ into $q_x$. One then removes it at the end of the estimates. Since such a procedure is standard, we omit it here.) This relation, together with 
\begin{align} \label{estPhi-qed}
|\lan \psi, \Phi ( i \om^{-1} q_x )\psi\ran|\ls \Big ( \int\om^{-3- 2\nu +\eps} \langle k \rangle^{-6} dk \Big )^{\frac12 } \big \| \d \Gamma( \om^{-\eps} )^{\frac12} \psi \big \| \| \psi \|,
\end{align} 
for any $\eps>0$, which follows from the bounds of Lemma I.1 of Supplement I, and
\begin{align} \label{omq-est-qed}
\|\om^{-\frac12} q_x \|_{\mathfrak{h}}\ls \|\om^{- 1 - \nu} \langle k \rangle^{-3} \|_{\mathfrak{h}},
\end{align}
implies \eqref{1overk-est-qed}.

To prove \eqref{dGy-est-qed} and \eqref{dGtildeb-est-qed}, we proceed similarly, using, instead of \eqref{estPhi-qed} and  \eqref{omq-est-qed},  the estimates 
\begin{align*}% \label{estPhi2-qed}
& |\lan \psi, \Phi ( i \lan y \ran q_x )\psi\ran| \ls \Big ( \int\om^{- 2 - 2 \nu } \lan k \ran^{-6}dk \Big )^{\frac12} \big \| \d \Gamma( \om^{-1} )^{\frac12} \psi \big \| \| \< x \>\psi \| \notag \\
& \phantom{ |\lan \psi, \Phi ( i \lan y \ran q_x )\psi\ran|}  \ls \Big ( \int\om^{-2- 2 \nu } \lan k \ran^{-6}dk \Big )^{\frac12} \big \| \d \Gamma( \langle y \rangle )^{\frac12} \psi \big \| \| \< x \>\psi \|, \\
& \|\lan y \ran^{\frac12} q_x \|_{\mathfrak{h}}\ls \< x \>^{\frac12} \|\om^{-1- \nu} \lan k \ran^{-3} \|_{\mathfrak{h}},  
\end{align*} 
and
\begin{align*} %\label{estPhi3-qed}
& \| \Phi ( i \tilde b q_x )\psi\| \ls \Big ( \int\om^{-2 - 2\nu} \lan k \ran^{-6} d k \Big )^{\frac12} \| \< x \> ( H_f +1 )^{\frac12} \psi \| , \\
& \big | \lan q_x, \tilde b q_x \ran_{\mathfrak{h}} \big | \ls \< x \>\|\om^{-\frac12 - \nu } \lan k \ran^{-3} \|_{\mathfrak{h}}^2. 
\end{align*}

\subsection{ Proof of Theorem  \ref{thm:ac}}\label{sec:ac-qed}
We  present the parts of the proof of Theorem  \ref{thm:ac} for the hamiltonian \eqref{Hqed-one-b} which differ from that  for the hamiltonian \eqref{H}--\eqref{I}. To begin with, the existence and the properties of  the Deift-Simon wave operators, 
\begin{equation*} %\label{W-def}
W_\pm := \slim_{t\to \pm \infty} W(t),\ \quad \mbox{with}\ \quad  W(t) := e^{ i t \hat{H}}\check{\Gamma} (j) e^{- i t H},
\end{equation*}
on $\mathrm{Ran}_{ ( - \infty , \Sigma ) }( H )$, where $ \hat H = H \otimes \one + \one \otimes H_f$ and the operators  $\check{\Gamma} $ and $j=(j_0, j_\infty)$ are defined in Subsection 5.1 of \cite{FauSig1}, are equivalent to the existence and the properties of the modified  Deift-Simon  wave operators
\begin{equation} \label{W-mod-def}
W_\pm^{(\rm mod)} := \slim_{t\to \pm \infty} \big( e^{- i \Phi (q_{x})} \otimes \one \big ) e^{ i t \hat{H} } \check{\Gamma} (j) e^{ - i t H} e^{i\Phi (q_{x})},
\end{equation}
on $\mathrm{Ran}_{ ( - \infty , \Sigma ) }( \tilde H )$ (where $\tilde H = e^{ - i \Phi( q_x ) } H e^{ i \Phi( q_x ) }$ is given in \eqref{tildeH}).

To prove the existence of $W_\pm^{(\rm mod)}$, we observe that, due to
\begin{align}\label{checkG-a}
\check{\Gamma}(j) a^\#(h) &= ( a^\#(j_0 h) \otimes \one + \one \otimes a^\#(j_\infty h) ) \check{\Gamma}(j),
\end{align}
where $a^\#$ stands for $a$ or $a^*$, we have $\check{\Gamma}(j) \Phi (h) = \hat\Phi (h) \check{\Gamma}(j)$, where
\begin{align*}%\label{hatPhi}
\hat\Phi (h) &:= \Phi (j_0 h) \otimes \one + \one \otimes \Phi (j_\infty h),
\end{align*}
which in turn implies that
\begin{align*}%\label{checkG-ePhi}
\check{\Gamma}(j) e^{i\Phi (h)} &= e^{i\hat\Phi (h)} \check{\Gamma}(j) .
\end{align*}
Therefore
\begin{align}\label{DS-transf}
\big( e^{- i \Phi (q_{x})} \otimes \one \big ) e^{ i t \hat{H} } \check{\Gamma} (j) e^{ - i t H} e^{i\Phi (q_{x})} &= \big( e^{- i \Phi (q_{x})} \otimes \one \big ) e^{ i t \hat{H} } e^{ i \hat \Phi( q_x ) } \check{\Gamma} (j) e^{- i t \tilde H  }  \notag\\
& = e^{ i t \hat H^{(\mathrm{mod})} } \check{\Gamma} (j) e^{- i t \tilde H } + \mathrm{Rem}_t ,
\end{align}
where $\hat H^{(\mathrm{mod})} := \tilde H \otimes \one + \one \otimes H_f$ and $\mathrm{Rem}_t := \big( e^{- i \Phi (q_{x})} \otimes \one \big ) e^{ i t \hat{H} } \big ( e^{ i \hat \Phi( q_x ) } - e^{ i \Phi (q_{x})} \otimes \one\big ) \check{\Gamma} (j) e^{- i t \tilde H }$. We claim that
\begin{align}\label{Rt-decay}
\slim_{t \to \pm \infty} \mathrm{Rem}_t = 0.
\end{align}
Indeed, let $R := \hat \Phi( q_ x ) - \Phi( q_x ) \otimes \one =  \Phi( ( j_0 - 1 ) q_x ) \otimes \one + \one \otimes \Phi( j_\infty q_x )$ and $\hat N := N \otimes \one + \one \otimes N$. Using \eqref{<}, Lemma I.1 and Lemma B.6 of \cite{FauSig1} (see also \cite[Lemma 3.1]{BoFaSig}), we obtain
\begin{align*}
& \big \|ÊR( \hat N + 1 )^{-\frac12} \big \| \lesssim \| ( j_0 - 1 ) q_x \|_{ \mathfrak{h} } + \| j_\infty q_x \|_{ \mathfrak{h} } \lesssim t^{ - \alpha \tau} \langle x \rangle^{1+\tau}, 
\end{align*}
for any $\tau < 1$. From this estimate and the relation $e^{ i \hat \Phi( q_x ) } - e^{ i \Phi( q_x ) } \otimes \one=-i \int_0^1ds e^{(1-s) i \hat \Phi( q_x ) } R ( e^{s i \Phi( q_x ) } \otimes \one)$, it is not difficult to deduce that 
\begin{equation*}
\big \|Ê\big ( e^{ i \hat \Phi( q_x ) } - e^{ i \Phi( q_x ) } \otimes \one \big ) ( \hat N + \langle x \rangle^{2+2\tau} + 1 )^{-1} \big \| \lesssim t^{ - \alpha \tau} .
\end{equation*}
Furthermore, we have $( \hat N + \langle x \rangle^{2+2\tau} + 1 ) \check{\Gamma}(j) = \check{\Gamma}(j) ( N + \langle x \rangle^{2+2\tau} + 1 )$, and, as in Corollary A.3 of \cite{FauSig1}, one can verify that $\| N e^{ - i t \tilde H } \psi_0 \| \lesssim t^{2/5} \| \psi_0 \|_{-1}$ for any $\psi_0 \in f( \tilde H ) D( K^{1/2} )$, $f \in \mathrm{C}_0^\infty ( ( - \infty , \Sigma ) )$. Using in addition that $\| \langle x \rangle^{2+2\tau} f( \tilde H ) \| < \infty$, it follows that $\mathrm{Rem}_t$ strongly converges to 0 on $\mathrm{Ran}_{ ( - \infty , \Sigma ) }( \tilde H )$ provided that $\alpha \tau > 2/5$.  

The equations \eqref{W-mod-def}, \eqref{DS-transf} and \eqref{Rt-decay} imply
\begin{equation}\label{DSmod-expr}
 W_\pm^{(\rm mod)} = \slim_{t \to \pm \infty} e^{ i t \hat H^{(\mathrm{mod})} } \check{\Gamma} (j) e^{- i t \tilde H } .
\end{equation}
The proofs of the existence and properties of the Deift-Simon wave operators \eqref{DSmod-expr} are then done as for $ W_\pm$ in \cite{FauSig1} (see Theorem 5.1 of \cite{FauSig1}),  but with $G_1$, defined in (5.16) of \cite{FauSig1}, replaced by
\begin{equation}\label{tildeG1}
\tilde G_1 := (\tilde I ( g) \otimes \bfone ) \check{\Gamma} (j) - \check{\Gamma} (j) \tilde I ( g),
\end{equation}
where $\tilde I ( g):=p \cdot \tilde{A} (x) + \tilde{A} (x) \cdot p + \tilde{A} (x)^2 + E(x)$ is coming from the decomposition
\begin{equation*}%\label{tildeH-deco}
\widetilde{H}  = H_p + H_f + \tilde I ( g).
\end{equation*}
We use \eqref{checkG-a} to derive 
\begin{align*}%\label{checkG-tildeI(g)}
\check{\Gamma}(j) \tilde I ( g ) = \hat I ( g )\check{\Gamma} ( j ),
\end{align*}
where 
\begin{align*}%\label{hatI} 
\hat I ( g ) :=  p \cdot \hat{A} (x) + \hat{A} (x) \cdot p + \hat{A} (x) ^2 + \hat{E} (x) ,
\end{align*}
with $\hat{A} (x):= \Phi( j_0\tilde{g}_{x} ) \otimes \one + \one \otimes  \Phi( j_\infty \tilde{g}_{x} )$, $\hat{E} (x):= \Phi( j_0e_{x} ) \otimes \one + \one \otimes  \Phi( j_\infty e_{x}  ) $.  This together with \eqref{tildeG1} gives
\begin{align*}%\label{tildeG1-expr}
 \tilde G_1 =  I_\#(  g ) \check{\Gamma}(j) ,
\end{align*} 
where 
\begin{align*}%\label{sharpI}  
I_\#(  g ):=  p \cdot A_\# (x)+ A_\# (x) \cdot p + A_\#  (x)\cdot \hat{A} (x) + \hat{A} (x)\cdot A_\#  (x) + E_\#  (x) ,
\end{align*}
with  $A_\#  (x):= \Phi((1- j_0)\tilde{g}_{x} )\otimes \one - \one \otimes   \Phi( j_\infty\tilde{g}_{x} ) $, $E_\#  (x):= \Phi( (1- j_0)e_{x} ) \otimes \one - \one \otimes  \Phi( j_\infty e_{x}  )  $.  Using the last expression, Lemma I.1 and the estimates 
\begin{align*}%\label{JinftyJ0-est}
& \| j_\infty  \tilde g_x \|_{\mathfrak{h}} \lesssim t^{ - \alpha \tau } \langle x \rangle^{ \frac{1}{\nu} + \tau }, \quad \| (1 - j_0) \tilde g_x \|_{ \mathfrak{h} }\lesssim t^{ - \alpha \tau } \langle x \rangle^{ \frac{1}{\nu} + \tau }, \\
& \| j_\infty  e_x \|_{\mathfrak{h}} \lesssim t^{ - \alpha \tau } \langle x \rangle^{ 1 + \tau }, \quad \| (1 - j_0) e_x \|_{ \mathfrak{h} }\lesssim t^{ - \alpha \tau } \langle x \rangle^{ 1 + \tau },
\end{align*}
with $\tau<2$ (see Lemma B.6 of \cite{FauSig1}), we arrive at the following refinement of estimate (5.31)  of \cite{FauSig1}
\begin{align*}
\| f( \hat H ) G_1 ( N + 1)^{-1/2} \| \lesssim t^{-\alpha \tau} . 
\end{align*}
The remainder of the proof goes the same way as the proof of Theorem 5.1 of \cite{FauSig1}. Hence $W_\pm^{(\rm mod)}$ and therefore $W_\pm$ exist.

\medskip

Now, we comment on the proof of the key Theorem 5.4 of \cite{FauSig1} in the QED case. It goes in the same way as for the phonon model \eqref{H}--\eqref{g-est} (see \cite{FauSig1}), until the point where we have to show that $\| \d \Gamma( b_\e^2 )^{1/2} P_{ \mathrm{gs} } \| = \mathcal{O}( t^\kappa )$ in the present case. This estimate can be proven by using the generalized Pauli-Fierz transformation \eqref{gen-PF} together with \eqref{Phi6}, to obtain
\begin{equation}\label{eq:est-dbe}
\big \| \d \Gamma( b_\e^2 )^{\frac12} \Phi_{ \mathrm{gs} } \big \|^2 = \Big \langle \tilde \Phi_{ \mathrm{gs} } ,  \big ( \d \Gamma( b_\e^2 ) - \Phi ( i b_\e^2 q_x ) + \frac12 \langle b_\e^2 q_x , q_x \rangle_{ \mathfrak{h} } \big ) \tilde \Phi_{ \mathrm{gs} } \Big \rangle ,
\end{equation}
where $\tilde \Phi_{ \mathrm{gs} } := \mathcal{U} \Phi_{ \mathrm{gs} }$. Using Lemma I.1 of Supplement I, \eqref{exp-bnd} and the fact that $\tilde \Phi_{ \mathrm{gs} } \in D( N^{1/2} )$, we can estimate the second term of the r.h.s. of \eqref{eq:est-dbe} as
\begin{equation*}
\Big | \Big \langle \tilde \Phi_{ \mathrm{gs} } ,   \Phi ( i b_\e^2 q_x ) \tilde \Phi_{ \mathrm{gs} } \Big \rangle \Big | \le \big \| \langle x \rangle^{3} \tilde \Phi_{ \mathrm{gs} } \big \| \big \| \langle x \rangle^{-3} \Phi( i b_\e^2 q_x ) ( N+1 )^{-\frac12} \big \| \| (N+1)^{\frac12} \tilde \Phi_{ \mathrm{gs} } \| \lesssim t^{2\kappa}.
\end{equation*}
Likewise, $| \langle \tilde \Phi_{ \mathrm{gs} } , \langle b_\e^2 q_x , q_x \rangle_{ \mathfrak{h} } \tilde \Phi_{ \mathrm{gs} } \rangle | \lesssim t^{2\kappa}$. To estimate the first term of the r.h.s. of \eqref{eq:est-dbe}, we write
\begin{align*}
\big \| \d \Gamma( b_\e^2 )^{\frac 12} \tilde \Phi_{ \mathrm{gs} } \big \|^2 = \sum_\lambda \int \big \| b_\e a_\lambda(k) \tilde \Phi_{ \mathrm{gs} } \big \|^2 dk.
\end{align*}
Applying the standard pull-through formula gives
\begin{align*}
a_\lambda(k) \tilde \Phi_{ \mathrm{gs} } = - \big ( \tilde H - E_{Ê\mathrm{gs} } + |k| \big )^{-1}Ê\big ( ( p + \tilde A ( x ) ) \cdot \tilde g_x(k) + e_x(k) \big ) \tilde \Phi_{ \mathrm{gs} }.
\end{align*}
We then easily conclude that $\| b_\e a_\lambda(k) \tilde \Phi_{ \mathrm{gs} } \|_{ \mathfrak{h} } = \mathcal{O}( t^\kappa )$ in the same way as in Corollary C.4 of \cite{FauSig1}.

\bigskip

\section*{Supplement I. Photon creation and annihilation operators}\label{sec:crannihoprs}

Recall that the propagation speed of the light and the Planck constant divided by $2 \pi$ are set equal to 1. We use the momentum representation and write functions from the space $\fh= L^2(\R^3, \C^2)$ as $u(k, \lam)$, where $k\in\R^3$ is the wave vector or momentum of the photon and $\lam\in \{-1, +1\}$ is its polarization.

With each function $f \in \fh$, one associates \emph{creation}  and \emph{annihilation operators} $a(f)$ and $a^*(f)$ defined, $\mbox{for}\  u\in \otimes_s^n\fh,$ as
\begin{equation*}%\label{cr-annih-ops}
a^*(f)  : u\ra \sqrt{n+1}  f\otimes_s u\ \quad   \mbox{and}\  \quad  a(f)  : u\ra \sqrt{n}  \lan f, u\ran_\fh, %\tag{I.1}
\end{equation*}
with $\lan f, u\ran_\fh :=\sum_{\lambda = 1 , 2} \int  d k \, \overline{f(k, \lam)} u_{n}(k, \lam, k_1, \lambda_1, \ldots, k_{n-1}, \lambda_{n-1})$. They are unbounded, densely defined operators of $\G(\fh)$, adjoint of each other (with respect to the natural scalar product in $\cF$) and satisfy the \emph{canonical commutation relations} (CCR):
\begin{equation*}
\big[ a^{\#}(f) , a^{\#}(g) \big] = 0 , \qquad \big[ a(f) , a^*(g) \big] = \lan f, g\ran ,
\end{equation*}
where $a^{\#}= a$ or $a^*$. Since $a(f)$ is anti-linear and $a^*(f)$ is linear in $f$, we write formally
\begin{equation*}
a(f) = \sum_{\lambda = 1 , 2} \int  d k \, \overline{f(k, \lam)} a_\lambda(k) , \qquad a^*(f) = \sum_{\lambda = 1 , 2} \int d k \, f(k, \lam) a_\lambda^*(k) ,
\end{equation*}
for photons. Here $a_\lambda(k)$ and $a_\lambda^*(k)$ are unbounded, operator-valued distributions, which obey (again formally) the \emph{canonical commutation relations}:
\begin{equation*}
\big[ a_{\lambda}^{\#}(k) , a_{\lambda'}^{\#}(k') \big] = 0 , \qquad \big[ a_{\lambda}(k) , a_{\lambda'}^*(k') \big] = \delta_{\lambda, \lambda'} \delta (k-k') ,
\end{equation*}
where $a_\lambda^{\#}= a_\lambda$ or $a_\lambda^*$.

Given an operator $\tau$ acting on the one-particle space $\fh$, the operator $\d \Gamma( \tau )$ defined on the Fock space $\cF$ by \eqref{dG} can be written (formally) as $\d \Gamma( \tau ) : = \sum_{\lambda = 1 , 2} \int d k \, a_\lambda^* ( k ) \tau a_\lambda ( k )$. Here the operator $\tau$ acts on the $(k, \lam)$-variable ($\d \Gamma( \tau )$ is the second quantization of $\tau$). The precise meaning of the latter expression is \eqref{dG}. In particular, one can rewrite the quantum Hamiltonian $H_f$ in terms of the creation and annihilation operators, $a$ and $a^*$, as
\begin{equation*}
H_f = \sum_{\lambda = 1 , 2} \int \, d k \, a_\lambda^*(k) \omega(k) a_\lambda(k).
\end{equation*}

Commutators of two  $\d \Gamma$ operators reduces to commutators of the one-particle operators:
\begin{align*}
[\d \Gamma ( \tau ) , \d \Gamma ( \tau' )]  = \d \Gamma ( [ \tau , \tau' ] ).  %\tag{I.2} \label{Phi3} 
\end{align*}

Let $\tau$ be a one-photon self-adjoint operator.   The following commutation relations involving the field operator $\Phi ( f ) = \frac{1}{\sqrt{2}} ( a^{*} ( f ) + a (f ))$ can be readily derived from the definitions of the operators involved:
\begin{align*}
& [ \Phi (f) , \Phi (g) ] = i \im \< f , g \>_{\mathfrak{h}} , \\% \label{Phi1}\\
& [ \Phi (f) , \d \Gamma ( \tau ) ] = i \Phi ( i \tau f ) , \\%  \label{Phi2} \\
& [{\Gamma}(\tau), \Phi (f)]=\Gamma( \tau ) a ((1-\tau) f) - a^*((1-\tau) f)\Gamma( \tau ) .  % \label{Phi6'}
 \end{align*}
Exponentiating these relations, we obtain 
\begin{align}
& e^{i \Phi (f)} \Phi (g) e^{- i \Phi (f)} = \Phi (g) - \im \< f , g \>_{\mathfrak{h}} ,  \tag{I.1} \label{Phi4}\\
& e^{i \Phi (f)} \d \Gamma ( \tau ) e^{- i \Phi (f)} = \d \Gamma ( \tau ) - \Phi ( i \tau f ) + \frac{1}{2} \re \< \tau f , f \>_{\mathfrak{h}} \tag{I.2}  \label{Phi5}\\
& e^{i \Phi (f)}  \Gamma ( \tau ) e^{- i \Phi (f)} =  \Gamma ( \tau )  + \int_0^1 ds \, e^{i s\Phi (f)}  ( \Gamma( \tau ) a ((1-\tau) f) - a^*((1-\tau) f)\Gamma( \tau ) ) e^{-s i \Phi (f)} . \tag{I.3}  \label{Phi6}
 \end{align}

Finally, we have the following standard estimates for annihilation and creation operators $a(f)$ and   $a^*(f)$, whose proof can be found, for instance, in \cite{BFS2}, \cite[Section 3]{GGM2}, \cite{GuSig}: \\

\noindent \textbf{Lemma I.1}   \textit{
For any $f \in \mathfrak{h}$ such that $\omega^{-\rho/2} f \in \mathfrak{h}$, the operators $a^\#(f) ( \d\Gamma( \omega^\rho ) +1 )^{-1/2}$, where $a^\#(f)$ stands for $a^{*}(f)$ or $a(f)$, extend to bounded operators on $\mathcal{H}$ satisfying
\begin{align*}
& \big\Vert a(f) ( \d\Gamma( \omega^\rho ) +1 )^{- \frac{1}{2} } \big\Vert \leq \Vert \omega^{-\rho/2} f \Vert_{ \mathfrak{h}} ,    \\
& \big\Vert a^{*} (f) (  \d\Gamma( \omega^\rho ) +1 )^{- \frac{1}{2} } \big\Vert \leq \Vert \omega^{-\rho/2} f \Vert_{ \mathfrak{h} } + \Vert f \Vert_{ \mathfrak{h} }.
\end{align*}
If, in addition, $g \in \mathfrak{h}$ is such that $\omega^{-\rho/2} g \in \mathfrak{h}$, the operators $a^\#(f) a^\#(g) ( \d\Gamma( \omega^\rho ) +1 )^{-1}$ extend to bounded operators on $\mathcal{H}$ satisfying
\begin{align*}
& \big\Vert a(f) a(g) ( \d\Gamma( \omega^\rho ) +1 )^{ -1 } \big \Vert \leq \Vert \omega^{-\rho/2} f \Vert_{ \mathfrak{h}} \Vert \omega^{-\rho/2} g \Vert_{ \mathfrak{h}} ,    \\
& \big\Vert a^*(f) a(g) ( \d\Gamma( \omega^\rho ) +1 )^{ -1 } \big \Vert \leq \big ( \Vert \omega^{-\rho/2} f \Vert_{ \mathfrak{h} } + \Vert f \Vert_{ \mathfrak{h} } \big ) \Vert \omega^{-\rho/2} g \Vert_{ \mathfrak{h}} ,    \\
& \big\Vert a^{*} (f) a^*(g) (  \d\Gamma( \omega^\rho ) +1 )^{ - 1 } \big\Vert \leq \big ( \Vert \omega^{-\rho/2} f \Vert_{ \mathfrak{h} } + \Vert f \Vert_{ \mathfrak{h} } \big ) \big ( \Vert \omega^{-\rho/2} g \Vert_{ \mathfrak{h} } + \Vert g \Vert_{ \mathfrak{h} } \big ).
\end{align*}}
%\end{lemma}
%
%

\end{document}